\documentclass[aps,prl,twocolumn,groupedaddress,showpacs]{revtex4}

\usepackage{graphicx}
\usepackage{subfigure}

\begin{document}


\title{Multiscale Kinetic Monte-Carlo for Simulating 
Epitaxial Growth}

\author{Jason P. DeVita}
\email[]{jdevita@umich.edu}
\affiliation{Physics Department, University of Michigan, Ann Arbor, 
Michigan, 48109-1040}
\author{Leonard M. Sander}
\email[]{lsander@umich.edu}
\affiliation{Physics Department, University of Michigan, Ann Arbor, 
Michigan, 48109-1040}
\author{Peter Smereka}
\email[]{psmereka@umich.edu}
\affiliation{Department of Mathematics, University of Michigan, Ann 
Arbor, Michigan, 48109-1043}

\date{\today}

\begin{abstract}

We present a fast Monte-Carlo algorithm for simulating epitaxial surface
growth, based on the continuous-time Monte-Carlo algorithm of Bortz,
Kalos and Lebowitz.  When simulating realistic growth regimes, much
computational time is consumed by the relatively fast dynamics of the
adatoms.  Continuum and continuum-discrete hybrid methods have been
developed to approach this issue; however in many situations, the
density of adatoms is too low to efficiently and accurately simulate as
a continuum.  To solve the problem of fast adatom dynamics, we allow
adatoms to take larger steps, effectively reducing the number of
transitions required.  We achieve nearly a factor of ten speed up, for
growth at moderate temperatures and large $D/F$.

\end{abstract}


\maketitle

\section{\label{sec:intro}Introduction}

Molecular beam epitaxy (MBE) is a popular technique for growing
materials, and is also an interesting example of a non-equilibrium
statistical process.  A simplified view of MBE, which captures many of
the salient features, is the following.  Atoms arrive at a crystal
surface at a rate $F$.  Once on the surface, the atoms diffuse with
diffusion constant $D$ until they attach to an existing island or
nucleate a new island.  Atoms detach from island edges with rate $w_n=D
\exp(-n\varepsilon)$ where $n$ is the number of in-plane nearest
neighbors and $\varepsilon$ is the ratio of effective inter-atomic bond
energy to the the thermal energy ($k_BT$).  Over time, adatoms nucleate
new islands, islands grow by absorbing adatoms, and the surface grows
layer by layer.

A common technique for modeling epitaxial growth is kinetic Monte-Carlo
(KMC).  In KMC, atoms are treated as individual particles which diffuse
across flat substrates by taking nearest neighbor hops at a rate $4D$.
Atoms which are attached to islands execute nearest neighbor hops with a
reduced rate ($w_n$).  By representing each atom individually, KMC
automatically incorporates internal noise.  In many cases, KMC is a very
effective tool for simulating epitaxial growth, as it retains its
accuracy and flexibility through a wide range of growth environments.
However, this dependable accuracy and flexibility sometimes comes at the
expense of computational speed.  For example, when the adatom density
becomes large (such as at high temperature) or when the adatom diffusion
rate is much greater than the island-atom detachment rate, a KMC
simulation spends much of its computational effort computing adatom
dynamics.  To compensate for the adatom bottleneck, we devised a new
algorithm for simulating epitaxial growth, which is faster than KMC,
while retaining KMC's accuracy and flexibility.  Our method is a KMC
approach, but we modify the microscopic dynamics of the adatoms to
increase efficiency, while retaining accurate meso- and
macroscopic dynamics.  The essential step in the technique is to allow
the adatoms to hop many steps at once.  We show that this modified
dynamics reproduces the true dynamics in the regime of interest.


\section{\label{sec:kmc}Kinetic Monte-Carlo}

The most basic KMC algorithm used to simulate epitaxial growth, which we
will refer to as rejection KMC, goes as follows.  Consider an $L\times
L$ grid, choose a grid point at random, and calculate the hopping rate
for the atom at that site ($w_n = \exp(-n\varepsilon)$), where $n$ is
the number of in-plane nearest neighbor bonds.  Since adatoms have the
fastest hopping rates, we let our time-step be the time it takes for an
adatom to hop. Thus $\Delta t = 1$, and the probability for any atom to
hop in one time-step is $w_n$.  We then choose a random number between
$0$ and $1$, and if that number is less than $w_n$, the atom executes a
nearest neighbor hop.  To complete one time-step, this is repeated $L^2$
times -- enough so that on average every particle is given one chance to
move.

This algorithm is simple enough to be useful for a wide variety of
problems in epitaxial growth, including heteroepitaxy.  However, it can
be quite slow.  Since the hopping rates differ exponentially, many moves
are outright rejected.  To avoid this bottleneck, we use the BKL
algorithm \cite{bkl}, which samples according to the appropriate rates
without rejections.  The system we're simulating has a very simple rate
structure.  There are only five possible transitions in basic
homoepitaxy, corresponding to the $0$ to $4$ possible in-plane
bonds.  Thus we can use the BKL algorithm -- often referred to
as continuous-time KMC.  We keep lists of atoms
with each bond count -- a list of adatoms, a list of singly bonded
atoms, and list of doubly bonded atoms, and so on.  One loop of the
algorithm involves exactly one transition.  The hopping rate for each
type of atom is $w_n = \exp(-n\varepsilon)$.  The total transition rate
for the system is then $W = \sum_{j=0}^{4}w_jN_j$ where $N_0$ is the
number of adatoms, $N_1$ the number of singly bonded atoms, and so on.
The mean waiting time for a transition to occur is $1/W$.  So each loop
of the algorithm advances time by $1/W$ (which varies, since $N_j$
varies).  One of the five types of atoms is then chosen with probability
\begin{equation} P_j = \frac{w_jN_j}{W} \end{equation} and a random
member of that set of atoms is hopped. In a system with a few fast
moving atoms and many slow moving ones, BKL runs much faster than
rejection KMC, and so has become the standard for homoepitaxy
simulations.  It is worth noting that the ideas presented in this paper
are also applicable to rejection KMC, even though we focus on BKL.


\section{\label{sec:adatom_diff}Adatom diffusion}

As mentioned above, there are five rates in our problem.  The fastest is
adatom hopping ($w_0=1$), and the next fastest is hopping of singly
bonded atoms (exponentially slower at $w_1=\exp(-\varepsilon)$).  So if
$\varepsilon = 5$ (about $600 K$ for Cu for example), then about 148
adatom diffusion events occur for every singly bonded edge-atom hop --
and about $22,000$ adatom events for every doubly-bonded edge atom hop.
Thus, in regimes where adatom diffusion plays a significant role, much
of the computation is spent calculating the adatom trajectories. 

Our goal is to reduce the disparity in the transition rates.  We do this
by altering what we consider to be an adatom event.  Normally, an adatom
transition involves one nearest neighbor hop; instead, we allow an
adatom to diffuse for $n_d$ steps.  In other words, when we choose an
adatom, we allow it to execute $n_d$ random nearest neighbor hops before
moving on to the next atom.  This new $n_d$-step adatom transition must
occur at a rate $w_0' = 1/n_d$.  The total transition rate is the
sum of the rates for all the possible transitions \begin{equation} W =
\sum_{j=0}^{4}N_jw_j = N_0 + \sum_{j=1}^{4}N_jw_j \end{equation} which
becomes \begin{equation} W' = \frac{N_0}{n_d} + \sum_{j=1}^{4}N_jw_j
\end{equation} Since the $w_j$s are Boltzmann factors, the total rate is
dominated by the adatom term (when the temperature is low enough).  Thus
$W'$ is often much smaller than $W$.

Though the total transition rate has been greatly reduced, the included
events have become more complex and time-consuming (an $n_d$-step random
walk versus a one-step nearest neighbor hop).  Some increase in
efficiency is realized through this rate reduction technique, but there
is more to be gained.  For the growth conditions considered within this
paper (high $D/F$), the step density is often quite low, and many
adatoms are far from an step.  Given an adatom on an open terrace (no
step edges or other adatoms around) the subsequent motion is known a
priori (in a probabilistic sense).  If it is known that an adatom is far
enough from any obstacles, then the adatom can ``complete'' an n-step
random walk in one step.  The difficulty lies in knowing how much room
the adatom has.  A KMC method that includes long steps
has been developed in 1-D \cite{falk}; however tracking distances in
2-dimensions is significantly more challenging.

What is needed is a way of knowing the distance to the nearest
attachment site (a site with a free in-plane bond).  Several methods
have been developed to solve this issue in various related problems,
such as diffusion limited aggregation (DLA) \cite{dla}.  A fast method
for doing this that has been suggested \cite{ball-brady} is to use a
hierarchy of course-grained grids, where the depth of the hierarchy is
variable, and determined by the distance from an edge.  This works
nicely in the context of DLA -- where the growth is irreversible, there
is no nucleation, and there is only one random walker at a time.  But
when there are multiple random walkers, detachment, edge diffusion and
nucleation/breakup of dimers, the map hierarchy becomes too cumbersome
to be efficient.  Methods have also been developed to directly compute
the distance to an edge \cite{sethian,sethian2,tsai}.  Such methods are
very fast when computing the distance globally; they are too slow,
however, when only a local update is needed.

Here, we take a different approach to examining a adatom's local
environment: we search locally.  An n-step random walk
on square grid potentially covers an area of $2n(n+1)+1$ (a diamond of
diagonal length $2n+1$).  In other words, an adatom executing a 25-step
random walk can end up -- or pass through -- any of 1301 sites (all the
sites 25 or fewer hops away).  To simulate a 25-step random walk, we can
compute (a priori) the adatom probability density over the 1301-site
diamond, and then choose the final position of the adatom by sampling
randomly from the probability distribution.  This method is only
accurate if the 1301-site diamond is void of obstructions (steps,
attachment sites, and other adatoms).  Otherwise there would be a sink
for the adatom probability density, and the distribution would differ
from the a prior computed distribution.  So to replace a 25-step random
walk with a single hop requires searching 1301 sites for obstructions,
which is computationally cumbersome.

A simple approximation can reduce the number of sites
to be searched.  Consider that the probability density is very low at
the perimeter of the 1301-site diamond.  In fact, for an n-step random
walk, approximately 98\% of the probability density is contained within
a circle of radius $2\sqrt{n}$.  As a first approximation, we can
simulate the 25-step random walk by searching an area of about 314
sites, and then sample from the truncated probability distribution. 
That would be faster than searching all 1301 sites, but still too slow. 
We could continue to truncate the distribution closer and closer to the
origin, but the approximation becomes less and less accurate. 

Ideally, we would sample from a distribution that is as
small as possible, yet one that provides for accurate dynamics.  We
start by recognizing that the probability of finding a particle at
$(i,j)$ after a 1-step random walk is
\begin{equation}\label{eqn:randomwalk}
p_{i,j}^{t+1} = \frac{1}{4} (p_{i-1,j}^{t} + p_{i+1,j}^{t} +
p_{i,j+1}^{t} + p_{i,j-1}^{t}) 
\end{equation}
which becomes the diffusion equation
\begin{equation}\label{eqn:diffusion}
\frac{\partial p}{\partial t} = \frac{1}{4}\frac{\partial^2 p} 
{\partial x^2}
\end{equation}
in the continuum limit.  Given the approximation
\begin{equation}
\frac{\partial ^2 p}{\partial x^2} \approx \frac{1}{m^2} ( p_{i+m,j} -
2p_{i,j} + p_{i-m,j} ).
\end{equation}
we can replace equation (\ref{eqn:randomwalk}) with 
\begin{equation}\label{eqn:longstep}
p_{i,j}^{t+m^2} = \frac{1}{4} (p_{i-m,j}^{t} + p_{i+m,j}^{t} +
p_{i,j+m}^{t} + p_{i,j-m}^{t}).
\end{equation}
and retain the same continuum limit.  Essentially, we are sampling from
a distribution which contains only four points -- $(0,\pm m), (\pm
m,0)$, by allowing an adatom to hop a distance m in one of four 
directions.  To make sure that the long-hopping adatom doesn't skip over
anything, we need to search the square that extends out (m-1)-spaces --
$(2m-1)^2$ sites.  To simulate a 25-step random walk, we search a square
containing 81 sites, and then hop 5 spaces in one of 4 directions.  
Since each hop of a random walk requires the generation of a 
pseudo-random number, an 81-site search can be done faster than a 
25-step random walk.

Our algorithm is like the standard continuous time algorithm, but with
the following changes.  A step distance ($m$) is chosen,
so that the associated time step ($m^2$) does not
exceed the ratio of the adatom hopping rate to the next fastest process
(usually singly-bonded atoms).  All the rates $w_1 \rightarrow w_4$ are
unchanged, but the adatom transition rate $w_0$ becomes $1/m^2$.  Then,
if an adatom is chosen, we search the box around the adatom to find out
how far it can hop without hitting something (up to a maximum of $m$). 
If it has room to hop a distance $m$, then it does, and we move on the
the next transition.  If it has room only for a hop of distance $s < m$,
then it does so, but still has $m^2-s^2$ time left.  So it repeats the
process -- this time with a maximum step size of $\sqrt{m^2-s^2}$ --
until it uses up the full $m^2$ time-step.  That way all the adatoms
diffuse for the same length of time, but the ones near the edges (or
other adatoms) take several smaller steps instead of one large one.

Essentially we have replaced the typical adatom transition (one nearest
neighbor hop) with a rescaled adatom transition ($m^2$ hops), so that
the adatoms execute transitions with a rate nearer to that of singly
bonded atoms.  Furthermore, we can do the same to the one-bond atoms,
since they make transitions at a rate much faster than the doubly bonded
atoms.  The general idea is to have each type of atom operate on its own
time-scale, so that the rates become equalized.  Of course, for this
specific problem, there is little reason to go beyond singly bonded
atoms, so we stop there.  It is worth noting that if we were to have
enhanced edge-diffusion, this multiscale approach would provide
additional speed-up by also rescaling edge diffusion.


\section{\label{sec:alg}The algorithm}
We now assemble the ideas presented above into an algorithm for growing
on a substrate.  As in continuous-time KMC, each loop of the
algorithm is responsible for making one transition.  The time is then
updated according to the total transition rate, and the loop is run
until we reach the finishing time.

\begin{enumerate}

\item Compute the rates associated with each transition.  Let the number
of atoms with $k$ bonds be $N_k$; then the total transition rate is
\begin{equation} W = \frac{N_0}{m^2} +
\frac{N_1}{q}e^{-\varepsilon} + \sum_{k=2}^{4} N_ke^{-k\varepsilon}
+ F \end{equation} where $m$ is the hopping distance for an adatom (and
$m^2$ the rescaled time-step) and $q$ is the rescaled time-step for
singly bonded atoms.

\item Advance time by $1/W$.

\item Choose a type of atom to move.  Do this by generating a
pseudo-random number $r$ between $0$ and $W$.  If $r<F$ then we add
flux (one randomly placed atom); if $F\le r < N_0w_0$ do an adatom
event; and so on.  The six possible events are the five atom hops
(adatom, singly-bonded, etc.) and deposition.

\item Update the lists of each type of atom, and then restart the loop.

\end{enumerate}

\noindent Four of the six possible events are identical to
continuous-time KMC (adding flux, and hoping atoms with two, three or
four bonds).  The algorithm for an adatom event is as follows:

\begin{enumerate}

\item Choose uniformly from the list of adatoms.
\item Scan a square region of side-length $2m-1$.  Throughout the
simulation we keep track of the number of bonds available to attach to
at a given site.  Any site with more than zero attachment bonds is a
potential attachment site.  So, we start at the innermost square
surrounding the adatom, and sum up the attachment bonds in all the
sites in that ring.  If that sum is zero, we move on to the next
concentric square.  We keep doing that until either the sum is non-zero
or we reach the $(m-1)^{th}$ ring.  Then we know how long our step can
be, say $s$.
\item  We then take a step of size $s$ in a randomly chosen direction.
If $s = m$, then we are done; otherwise, we go back to step 2, but
with a maximum step of $\sqrt{m^2-s^2}$.  If at some point the
adatom attaches, we then choose another adatom to finish out the
time-step.

\end{enumerate}

\noindent The singly-bonded atoms act like the adatoms, except that
they only take nearest neighbor hops, and only for $q$ steps.


\section{\label{sec:results}Results}

\begin{figure}
\includegraphics[width=2.5in]{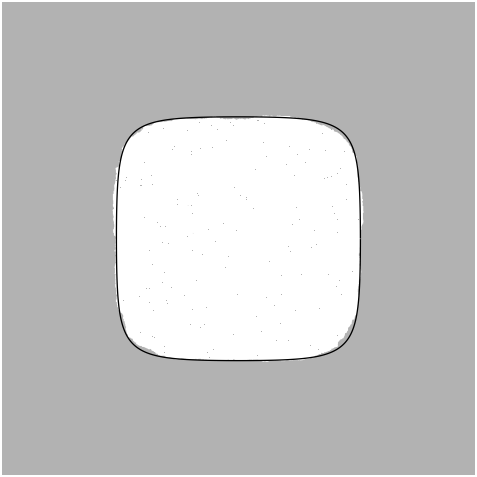}
\caption{\label{fig:eqis}
The equilibrium island shape for a large island (approximately
500,000 atoms) closely matches the exact (Wulff) solution (dark line).  
Because of the large size of the island, ensemble averaging is 
unnecessary.
The temperature is $E/kT = 4$.}
\end{figure}

To arrive at our MSKMC model, we've made some approximations regarding
the adatom dynamics.  While the approximations are physically
reasonable, it is important to verify that the results produced are
accurate.  Since we have only modified the dynamics, any equilibrium
results should be unaffected.  We can confirm this by computing
equilibrium island shapes.  The shape of an island in equilibrium with
the surrounding adatoms is known exactly by mapping onto the 2-d Ising
model \cite{wulff}.  We took as initial condition a square island on a
periodic grid, and allowed the island to come to equilibrium with the
surrounding adatoms.  The resulting island is shown in Fig.
\ref{fig:eqis} with the exact (Wulff) solution superimposed.  By using a
large island ($\sim 5\times 10^5$ atoms, or about $250$nm for copper),
we were able to achieve good agreement with the exact solution, without
ensemble averaging. This demonstrates that our method preserves the
original energetics.

\begin{figure}
\includegraphics[angle=0,width=3.3in]{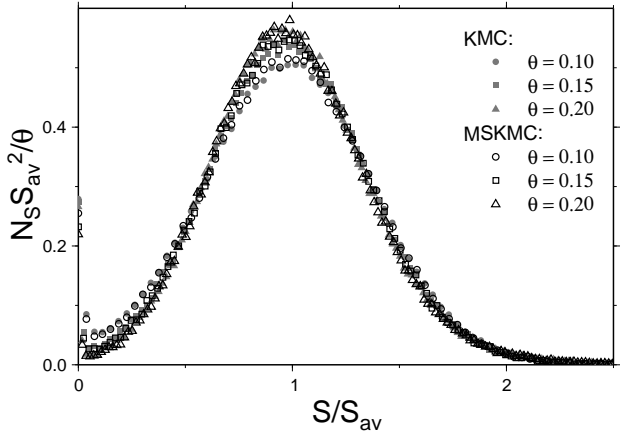}
\caption{\label{fig:isd_early}
Island size distribution for several coverages, showing good agreement
between KMC and MSKMC.}
\end{figure}

Since we have modified the dynamics of the adatoms, it is important to
verify that the new microscopic dynamics yield the correct meso- and
macroscopic results.  A common difficulty among continuum and
quasi-continuum methods is predicting the rate and location for
nucleation of new islands.  In models that treat that adatoms as a
continuum field, the formation of new islands must be computed from a
mean-field representation of the adatom density.  In situations where
fluctuations play a large role, a mean field treatment can give poor
results.  This is especially noticeable at early stages of growth, when
nucleation -- and subsequent breakup -- occur very rapidly, and
correlations in the adatom density can build up.  MSKMC avoids these
problems.  To demonstrate this, we compare it to ordinary
continuous-time Monte-Carlo.  Both models are allowed to grow starting
from a flat substrate, and the island sizes are tallied at various
coverages.  The results are shown in Fig. \ref{fig:isd_early}, where the
island size histograms are plotted for coverages of $10\%$, $15\%$, and
$20\%$.  There is good agreement between the two models.  This lends
confidence in the accuracy of our new adatom dynamics.  For this case,
the MSKMC simulation ran $6.5$ times faster than the standard BKL
simulation.

We now test the long-time kinetics, by observing mounding.  In the
presence of an Ehrlich-Schwoebel (ES) barrier, a growing surface
roughens and forms mounds.  The ES barrier is a step-edge barrier that
reduces the rate at which atoms ascend and descend steps.  This tends to
increase the adatom density on the island tops, thus increasing the
nucleation rate on top of existing islands.  This gives rise to a
mounding instability, which has been observed in both simulation and
experiment \cite{johnson,clarke,amar}.  The degree of mounding and mound
coarsening is sensitively dependent upon the nucleation on the island
tops \cite{krug-kuhn,krug-politi}, especially when the adatom density is
low.  This is one of the critical challenges of continuum based
modeling; mean-field derived nucleation rates are often inadequate for
predicting island-top nucleation \cite{castellano,krug-politi}.  The
sensitivity of mounding on island-top nucleation makes this a
challenging test for a method which modifies the adatom dynamics.

\begin{figure}
\includegraphics[width=3.3in]{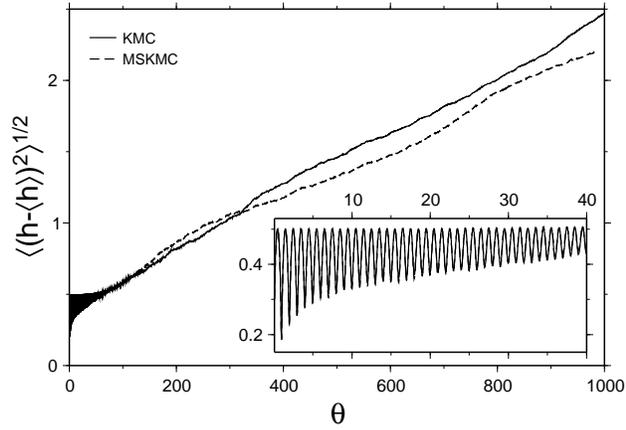}
\caption{\label{fig:es_rough}
When a small Ehrlich-Schwoebel barrier is included, the oscillations in 
the roughness diminish over time as the total roughness increases.  For 
this case, $E/kT = 5$, the grid is 1024 by 1024, $D/F = 10^6$, and the 
E-S barrier is $1kT$.  The difference between the two lines is a result 
of noise.}
\end{figure}

\begin{figure}
\mbox{
    \subfigure[KMC 500 ML]{
        \includegraphics[width=1.5in]{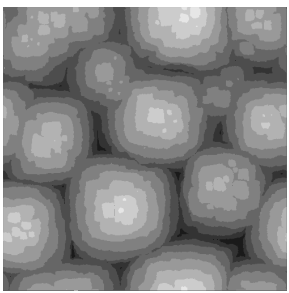}}
    \quad
    \subfigure[MSKMC 500 ML]{
        \includegraphics[width=1.5in]{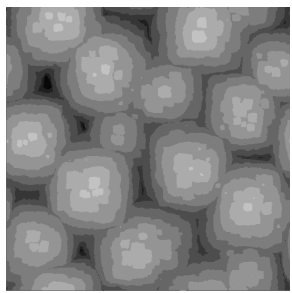}}
  }
\mbox{
    \subfigure[KMC 1000 ML]{
        \includegraphics[width=1.5in]{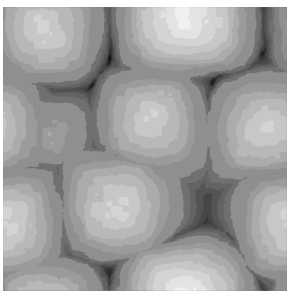}}
    \quad
    \subfigure[MSKMC 1000 ML]{
        \includegraphics[width=1.5in]{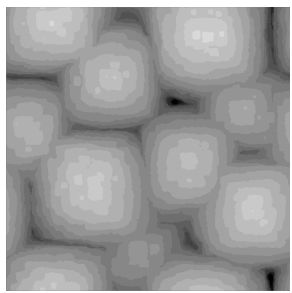}}
  }
\caption{\label{fig:es_mounds}
Surface images at 500, and 1000 monolayers, for
growth with a small Ehrlich-Schwoebel barrier.}
\end{figure}

Rejections are used to implement the ES barrier.  The
ES barrier can be handled using the BKL framework; however doing so
greatly increases the number of lists required, and thus the complexity
of the sorting.  Since the ES barrier is typically small and only arises
in certain configurations, the number of rejections is low -- and so
rejection can be an efficient alternative to sorting lists.  All the
other barriers in the simulation are handled using BKL, and when an ES
barrier is encountered, rejection is used.  We grew, using both regular
BKL and MSKMC, $1000$ monolayers on a $1024$ by $1024$ grid.  Figure
\ref{fig:es_rough} shows the surface roughness as a function of time. 
The roughness starts off oscillating; eventually, the oscillations die
down, and the overall roughness increases due to the mounding
instability.  The inefficiency of the BKL computation
makes ensemble averaging over multiple realizations prohibitive; however
the difference between the two results is consistent with observed
fluctuations.  Snapshots of the surface at $500$ and $1000$ monolayers
are shown in Fig. (\ref{fig:es_mounds}), for both KMC and MSKMC.  Our
MSKMC method ran $7$ times faster than the standard BKL method.  
Our method took about 24 hours to run the simulation on 
a 2.8GHz Pentium desktop.

\begin{figure}
\mbox{
    \includegraphics[width=3.3in]{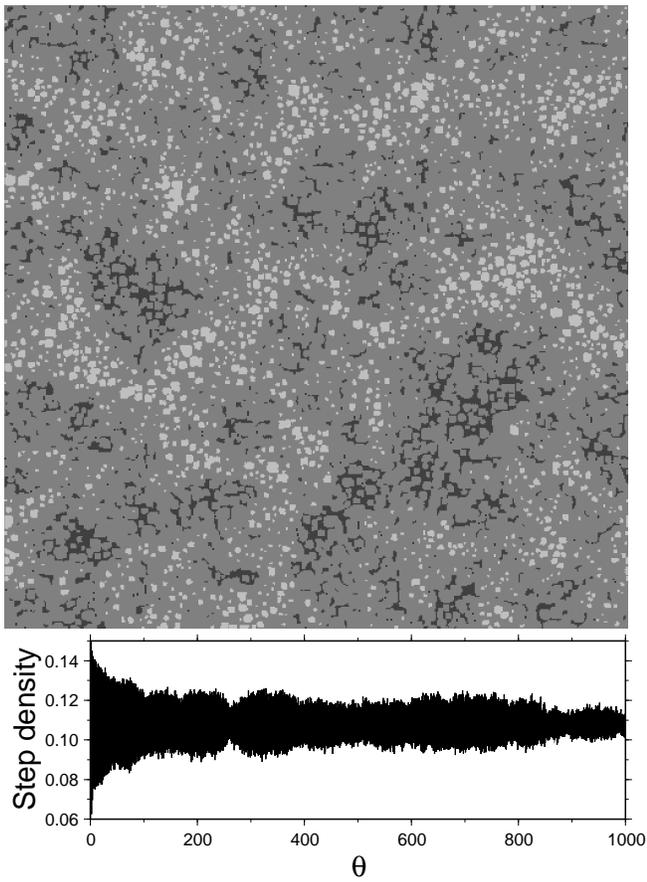}
}
\mbox{
    \includegraphics[width=3.3in]{noes_rough.ps}
}
\caption{\label{fig:noes}
Even in the absence of an ES barrier, the surface begins to roughen, 
and interesting morphology can arise.}
\end{figure}

Mounding can also occur in the absence of an ES barrier; however it can 
take a long time to happen.  Monte-Carlo simulations have shown 
this effect \cite{ratsch-wheeler}, but only when there is enhanced edge 
diffusion.  The increased efficiency of the MSKMC method allows us to 
simulate larger grid-sizes and longer times.   In Fig. \ref{fig:noes} we 
show the result of growing $1000$ monolayers on a $2048$ by $2048$ grid, 
with no ES barrier and no enhanced edge diffusion.  We are able to 
observe decay in the step-density oscillations, and also observe 
the formation of interesting large-scale features.  
This simulation took a little less than two days on a 
2.8GHz Pentium desktop.

Epitaxial growth on a vicinal surface is another problem of recent
interest.  In the presence of an ES barrier, step-flow becomes unstable
and the steps can meander considerably \cite{bales-zang}.  This effect
has been demonstrated computationally and experimentally
\cite{louis,kallunki,kato,hibino,maroutian}.  Some systems exhibit
meander wavelengths which are quite large.  For example, Hibino et. al. 
\cite{hibino} show experiments involving step-flow on Si, where the
meander wavelength is several microns.  Simulating such a system
requires a very large computational grid.  Coupled with the low
temperature and large $D/F$ typical of many of these experiments, KMC
can be quite slow.  Our MSKMC method is well suited to computations
involving large-scale features, low $T$ and high $D/F$.

\begin{figure}
  \subfigure[KMC]{
      \includegraphics[width=1.5in]{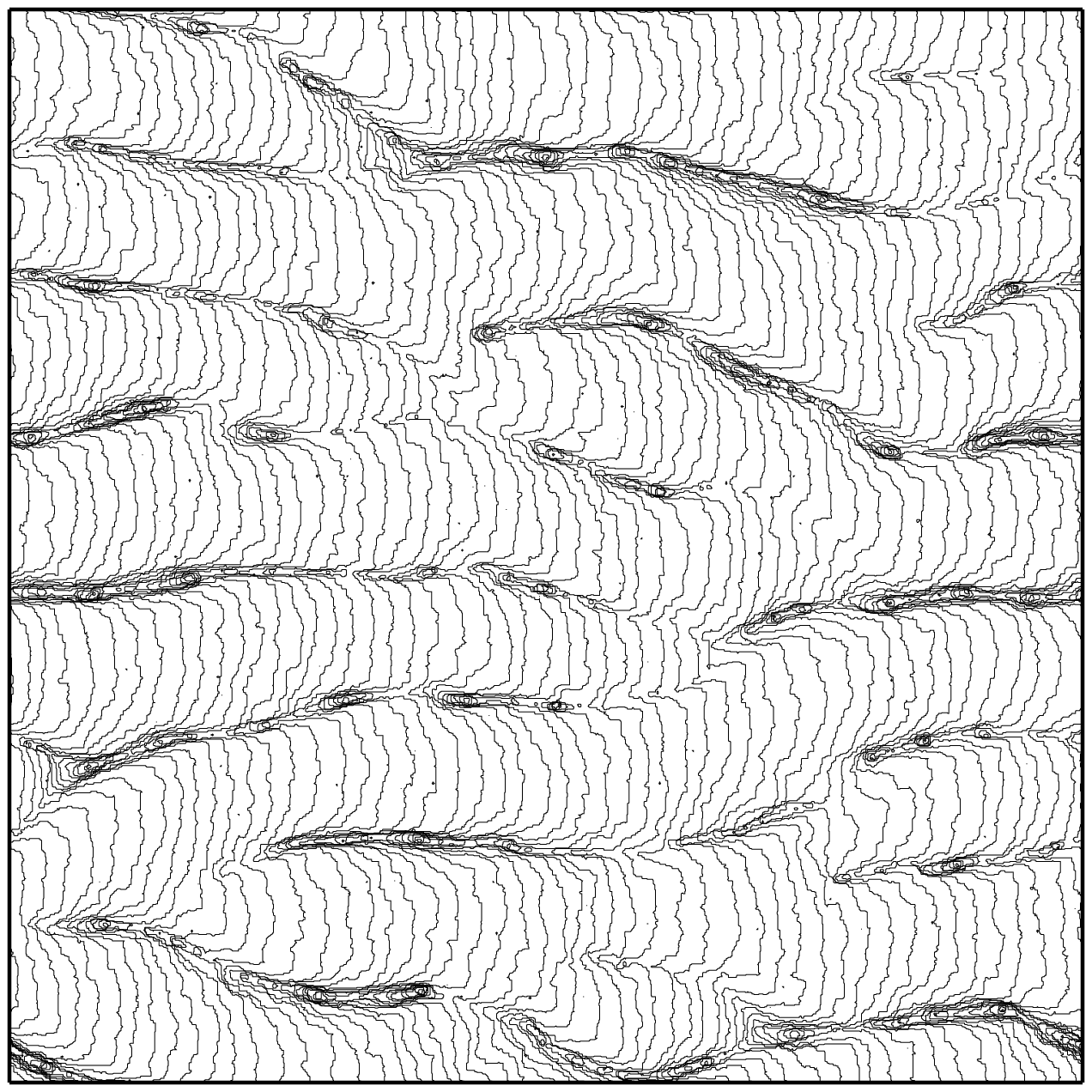}}
  \subfigure[MSKMC]{
      \includegraphics[width=1.5in]{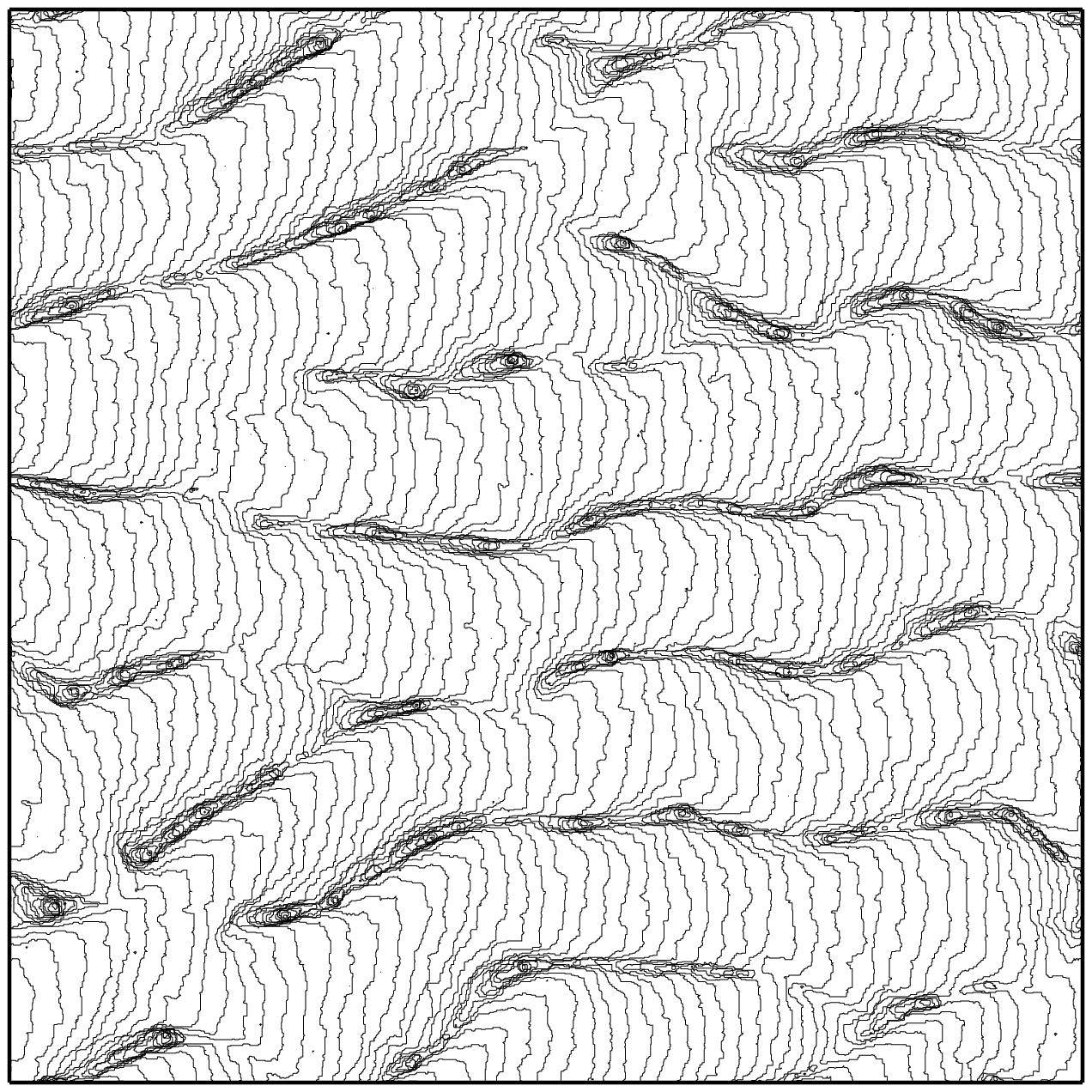}}
\caption{\label{fig:stepflow}
Starting from a uniform train of steps, a step-edge barrier causes the 
steps to meander.  The step spacing is 20 atoms, $E/kT = 5$, $E_{ES}/kT 
= 3$, and $D/F = 10^6$.}
\end{figure}

Figure \ref{fig:stepflow} shows and example of meandering step-flow. 
Both KMC and MSKMC are shown for comparison.  The computation was done
on a $1000$ by $1000$ grid, with an initial step spacing of $20$ atoms. 
The pattern that forms is very similar to patterns seen for growth on
vicinal Cu \cite{maroutian}.  The MSKMC method ran $9$ times faster than
the standard BKL method.  For smaller miscut (large step-spacing) and
lower temperature, the speed advantage would be even greater.

\section{\label{sec:disc}Discussion}

There have been several distinct attempts to increase the computational
efficiency of epitaxy simulation.  Here we will discuss a couple of
these other approaches.

\subsection{Absorbing Markov chains}

There are some similarities between our MSKMC
method and Novotny's MCAMC method \cite{novotny}.  Both methods desire
to improve computational efficiency by rescaling the fast dynamics;
however there are significant differences.  MCAMC follows the exact
dynamics in manner that waits for significant changes in the current
system state before executing a transition.  It requires knowledge of
the transition rates for not only the current state, but also for a
number of accessible states.  There are some situations for which such
knowledge is efficiently obtained, and as Novotny demonstrates,
significant speed-up can be seen.  Unfortunately, for a general epitaxy
simulation, it is simply too expensive to compute the transition rates
for nearby states.  It is therefore necessary to carefully approximate
the microscopic kinetics in a way that maintains the correct meso- and
macroscopic dynamics -- which is what we have done.

\subsection{Continuum models} Continuum models, which treat the adatoms
as a continuous fluid, have been suggested as an alternative to KMC
\cite{petersen,ratsch}.  Since the adatoms are represented by a
continuous density, the computational speed is independent of the number
of adatoms involved.  Continuum models with
deterministic attachment and detachment have had substantial success,
but their usefulness is limited to growth situations where fluctuations
are unimportant.  One key source of fluctuations is shot noise.  Shot
noise arises from the fact that atoms attach one at a time, as whole
particles.  Since diffusion-controlled growth is inherently unstable,
noise can play a vital role in determining the shape of islands.  An
extreme example is the diffusion limited aggregation (DLA) model of
Witten and Sander \cite{dla}.  DLA is a diffusive growth model with no
smoothing processes (such as edge diffusion or detachment), which
creates branching fractal structures.  Continuum models lack sufficient
noise to generate DLA-like clusters.

There has been some effort to include shot noise fluctuations in a
continuum model \cite{adkmc,schulze,schulze2,qcmc,qcmc2}.  
Most of
these methods utilize a hybrid approach where adatoms are modeled as a
continuous density, but islands are composed of discrete atoms.  Once
such approach is quasi-continuum Monte-Carlo (QCMC) \cite{qcmc,qcmc2}.
In QCMC, atoms are added to island edges as discrete particles, with
location determined (probabilistically) by the local adatom flux into
the island edge.  This incorporates shot noise fluctuations into the
continuum formalism.  This approach has had many successes, including
the ability to grow DLA-like clusters and reproduce correct thermal
roughening results \cite{qcmc,qcmc2}, and to some extent model
multilayer growth with mounding \cite{qcmc2}.

However, these hybrid models still have some shortcomings.  For one,
they're not necessarily faster than KMC.  In a continuum model, the
adatom density is defined over the entire grid, even if there are only a
few adatoms.  If the adatom density is low, then a continuum method must
take very long time steps when solving the diffusion equation in order
to be computationally efficient.  This can introduce error into the
computation.  Thus most continuum-based models are inefficient at low
adatom densities.  It is possible for such method to be efficient at
high densities; however, even that is challenging.  Since the nucleation
(and subsequent breakup) of new islands occurs at a higher rate as the
adatom density increases, the allowable time-step becomes shorter as the
density goes up.  It is a very delicate matter to take a sufficiently
long time step, while still obtaining acceptable results for nucleation.

Continuum-based methods bring difficulties beyond efficiency.  Since the
adatoms are replaced by a density, new islands must be explicitly
created.  Often that is done using a mean-field estimate of the
nucleation rate as a function of density.  There is a significant
literature on the subject of nucleation in epitaxy (see reference
\cite{ratsch-ven} for a review); much of which
uses a rate equation approach to calculate the dimer formation rate as a
function of density, temperature, and flux.  The results of such
calculations are used by many continuum-based models.  This appears to
work reasonably well in the absence of any step-edge barriers.  However,
it has been shown \cite{castellano,krug-politi} that when there is a
step-edge barrier, the rate
equation approach fails to predict the nucleation rate on island tops.
Moreover, growth in the presence of a step-edge barrier can lead to a
mounding instability which is very sensitively dependent on the
island top nucleation rate.

Our aim in developing the MSKMC method is to efficiently model adatom 
processes (and other fast processes) without the problems associated 
with continuum methods.  Continuum-based methods are very useful in 
situations where fluctuations are less important, and when one wants to 
closely parallel analytical approaches; however, when noise and discrete 
effects become important, a fully discrete method is often better 
suited.


\section{Conclusions}

We have presented an algorithm for simulating epitaxial growth, which is
between five and ten times faster than continuous-time kinetic Monte
Carlo for some interesting physical regimes.  Our multi-scale kinetic
Monte-Carlo (MSKMC) algorithm is a continuous-time method, which
rescales the adatom dynamics in both space and time.  For many realistic
growth conditions, the adatoms consume most of the computational time. 
By allowing adatoms to take long steps, we reduce the transition rate of
the dominant process.

It is worth noting that the relative efficiency of MSKMC (as compared to
BKL) is dependent on the random number generator.  Our method replaces
random processes (random walks) by deterministic ones (local searching).
Due to the computational complexity involved in generating a
pseudo-random number, an efficient search is faster than a random walk.
To be fair, we chose to use the fastest reasonable-quality random number
generator that we were able to obtain (a variation of a lagged Fibonacci
generator)\cite{Newman}, which is much faster than 
many compiler default
generators. Were we to use a slower generator (perhaps to ensure better
randomness) then our method would be even faster than we've stated,
relative to BKL.

To verify the validity of MSKMC, we've computed equilibrium island
shapes and early-growth island size distributions.  Both of which match
expected results -- analytical and computational.  We were able to
reproduce correct equilibrium island shapes without resorting to
ensemble averaging.  Our method also produces step-edge barrier
induced mounding that is consistent with standard KMC.  Due to our
increased efficiency, we can carry out simulations on larger grids
and/or for longer times, and perhaps observe phenomena previously
unseen.


\begin{acknowledgments} PS acknowledges useful
discussions with Michael Falk. This work is supported by the National
Science Foundation.  JPD and PS were supported by NSF grant numbers
DMS-0207402 and DMS-0509124.  LMS and PS were
supported by NSF grant number DMS-0244419. \end{acknowledgments}

\bibliography{mskmc}

\end{document}